# Nonlinear relativistic mean-field theory studies on He isotopes

G. W. Fan[1,*], T. K. Dong[2], D. Nishimura[3]

1 (School Of Chemical Engineering, Anhui University of Science and Technology, Huainan 232001, China)
2 (Purple Mountain Observatory, Chinese Academy of Sciences, Nanjing 210008, China)
3 (Department of Physics, Tokyo University of Science, Chiba 278-8510, Japan)

**Abstract:** The ground state properties of He isotopes are studied in the nonlinear relativistic mean-field (RMF) theory with force parameters NL-SH and TM2. The modified Glauber model as a gatekeeper is introduced to check the calculations. The investigation shows that the RMF theory provides a good description on the properties of He isotopes. The many-body space information of $^4$He + neutrons are obtained reliably. As a product, the calculation gives a strong evidence for neutron halo in $^5$He.

The $^7$He unbound ground state has been observed more than 40 years [1]; however, the positions and quantum numbers of excitations have remained uncertain. The excited states of $^7$He can decay not only to $^6$He + n channels, but also to $^5$He + 2n and $^4$He + 3n channels. This broad resonance condition makes it difficult to separate the excitations. Contradictions remain in the experiments, such as energy levels with spins of 5/2$^-$ and 1/2$^-$ [2-8]. Theoretically, ab initio calculations are often employed to describe the many-body decay properties of $^7$He; however, it is still unable to satisfy the multiparticle decay conditions for all open channels, because when one discusses the structure of $^7$He resonances, it is simultaneously important to describe the structure of $^5$He and $^6$He [9-16].

The nonlinear relativistic mean-field (RMF) theory has produced very reliable results on nuclear structure during past 30 years [17-20]. Therefore in this article we will calculate the ground state properties of $^{5,\,6,\,7}$He by use of the theory. Because the RMF theory is a standard theory and detailed formalism can be found in a number of Refs. [17-23], here we briefly describe the framework of the theory. In the RMF, the starting point for the description of the many-body problem is the local Lagrangian density

$$\mathcal{L} = \overline{\Psi}(i\gamma^\mu \partial_\mu - M)\Psi - g_\sigma \sigma \Psi - g_\omega \overline{\Psi}\gamma^\mu \omega_\mu \Psi$$
$$- g_\rho \overline{\Psi}\gamma^\mu \rho_\mu^a \tau^a \Psi + \frac{1}{2}\partial^\mu \sigma \partial_\mu \sigma - \frac{1}{2}m_\sigma^2 \sigma^2 - \frac{1}{3}g_2 \sigma^3$$
$$- \frac{1}{4}g_3 \sigma^4 - \frac{1}{4}\Omega^{\mu\nu}\Omega_{\mu\nu} + \frac{1}{2}m_\omega^2 \omega^\mu \omega_\mu - \frac{1}{4}R^{a\mu\nu}$$
$$\times R_{\mu\nu}^a + \frac{1}{2}m_\rho^2 \rho^{a\mu} \times \rho_\mu^a - \frac{1}{4}F^{\mu\nu}F_{\mu\nu} - \frac{1}{2}e\overline{\Psi}\gamma^\mu A^\mu(1-\tau^3)\Psi \qquad (1)$$

with relations

$$\Omega^{\mu\nu} = \partial^\mu \omega^\nu - \partial^\nu \omega^\mu, \qquad (2)$$
$$R^{a\mu\nu} = \partial^\mu \rho^{a\nu} - \partial^\nu \rho^{a\mu} + g_\rho \epsilon^{abc} \rho^{b\mu} \rho^{c\nu}, \qquad (3)$$

$$F^{\mu\nu} = \partial^\mu A^\nu - \partial^\nu A^\mu, \tag{4}$$

where $\rho_\mu^a$, $\sigma$, and $\omega_\mu$ denote the meson fields, $m$ with subscripts $\rho$, $\sigma$, and $\omega$ denotes their masses, respectively. The $\Psi$ and $M$ represent the nucleon field and rest mass. The $A^\mu$ is the photon field, which is responsible for the electromagnetic interaction $e^2/4\pi = 1/137$. The $g$ with subscripts 2, 3, $\rho$, $\sigma$, and $\omega$ represents the strengths of the coupling. The Pauli matrices are given by $\tau$. In practice the above parameters such as meson masses and coupling strengths are obtained through the fitting of the experimental observables. We will carry out numerical calculations with two sets of force parameters: NL-SH [23] and TM2 [24]. These two sets of force parameters are proposed by the fitting of properties ranging from light to heavy nuclei, even unstable nuclei.

**TABLE I.** The RMF results with NL-SH.

|  | $^5$He | $^6$He | $^7$He | $^8$He |
|---|---|---|---|---|
| B(MeV) | 33.21 | 33.43 | 34.82 | 37.34 |
| R(fm) | 2.60 | 2.68 | 2.71 | 2.74 |
| $R_p$(fm) | 1.88 | 1.86 | 1.84 | 1.83 |
| $R_n$(fm) | 2.80 | 2.84 | 2.83 | 2.83 |
| $R^2$(fm$^2$) | 16.52 | 12.81 | 11.32 | 10.55 |
| $-\epsilon(1s_{1/2})(p)$ | 20.32 | 24.54 | 28.86 | 33.14 |
| $-\epsilon(1s_{1/2})(n)$ | 19.42 | 20.85 | 22.21 | 23.48 |
| $-\epsilon(1p_{3/2})(n)$ | 0.81 | 1.77 | 2.74 | 3.68 |

**TABLE II.** The RMF results with TM2.

|  | $^5$He | $^6$He | $^7$He | $^8$He |
|---|---|---|---|---|
| B(MeV) | 33.52 | 34.07 | 35.74 | 38.39 |
| R(fm) | 2.53 | 2.65 | 2.71 | 2.76 |
| $R_p$(fm) | 1.88 | 1.86 | 1.84 | 1.83 |
| $R_n$(fm) | 2.69 | 2.80 | 2.83 | 2.85 |
| $R^2$(fm$^2$) | 14.75 | 12.42 | 11.35 | 10.76 |
| $-\epsilon(1s_{1/2})(p)$ | 21.51 | 25.94 | 30.40 | 34.74 |
| $-\epsilon(1s_{1/2})(n)$ | 20.28 | 21.71 | 23.04 | 24.24 |
| $-\epsilon(1p_{3/2})(n)$ | 1.15 | 2.05 | 2.94 | 3.78 |

The numerical results of nuclei $^{5,6,7,8}$He are listed in Tables I and II, where B (MeV), $R_m$ (fm), $R_p$ (fm), and $R_n$ (fm) are the binding energy, root-mean-square (RMS) radii of matter, proton, and neutron distributions. In order to explain the structure of He isotopes more fully, we also list the single particle energy $-\epsilon$ (MeV), and the mean-square radius of $1p_{3/2}$ neutron R$^2$ (fm$^2$), which reflects the many-body space information of $^4$He + X neutrons (Xn, X = 1, 2, 3.). It is seen that the calculations with NL-SH and TM2 force parameters exhibit almost the same results in two Tables. As shown by the single particle energy one can find that $1p_{3/2}$ neutrons are very weakly bound in He isotopes and $^{5,6}$He exhibits obviously as the neutron halo structure. It also reaches the same conclusion from the R$^2$, because R$^2$(1$p_{3/2}$)(n) = 15.64 ± 0.89 fm and = 12.62 ± 0.20 fm in $^5$He and $^6$He are large as compared with the mean-square radius of all neutrons $R_n^2 \approx$ 7.80 fm. The values of R$^2$(1$p_{3/2}$)(n) are the average values of the calculations listed in Table I and II, the errors are obtained from the difference between the calculations with NL-SH and TM2 force parameters. Experimentally, the $^6$He has been already proved to be a halo structure [25, 26]. In the Tables I and II, we also show the calculations of $^8$He, the small R$^2$, as compared with that of $^5$He and $^6$He, indicates that $^8$He has a skin structure, which is as good as conclusions of the experiments [26].

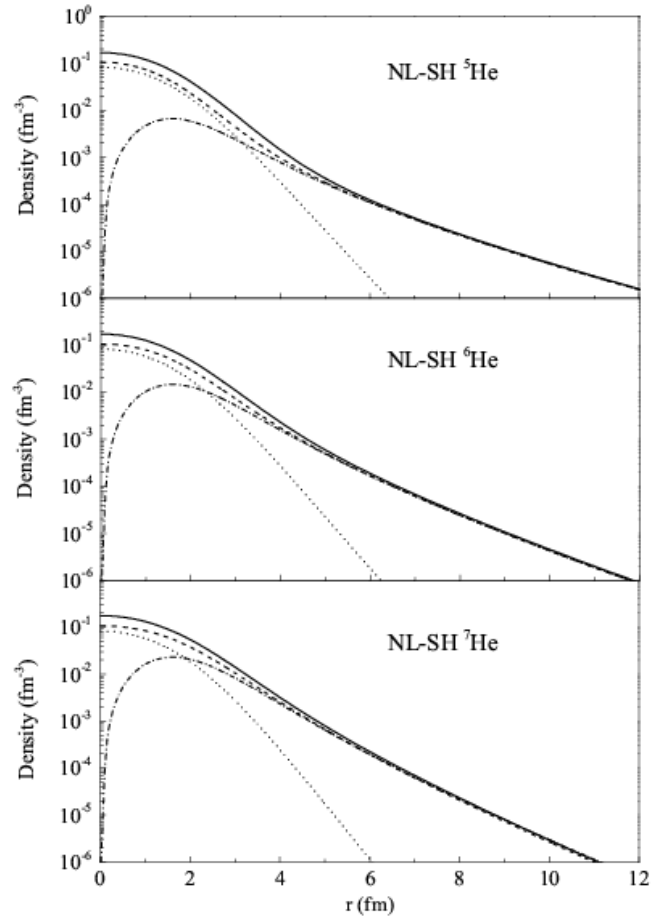

Fig. 1. The The density distributions of proton, neutron, matter, and Xn of He isotopes in the RMF theory with TM2 force parameters. Solid, dashed, dotted, and dash-dotted curves are matter, neutron, proton, and Xn, respectively.

In Fig. 1, we draw the density distributions of proton, neutron, matter and Xn in $^5$He, $^6$He and $^7$He. In the Fig., solid, dashed, dotted and dash-dotted curves are the densities of matter, neutron, proton, and Xn, respectively. It is evident that the halo structures in $^5$He and $^6$He as their density distribution of neutrons have a long tail. But ones also notice that due to the ambiguity of the definition of the halo structure, we can't conclude whether or not $^7$He is a halo. In order to deeply check the reliabilities of the results, we introduce the modified optical Glauber model [27] and the experimental reaction cross sections as gatekeepers. The Glauber model is often used to analyze the experimental reaction cross sections. It is a useful tool to extract the density distribution of the nuclei from the reaction cross sections. Thus the Glauber model is also an effective tool to check the reliabilities of the density distribution. In the calculations of the reaction cross sections of $^{6,8}$He + $^{12}$C, $^9$Be, and $^{27}$Al, the density distributions of the RMF theory with NL-SH and TM2 force parameters are used as the input parameters in the Glauber model, and the theoretical reaction cross sections are only a little difference between NL-SH and TM2. The outcomes with NL-SH parameter are shown in Fig. 2. It is concluded from Fig. 2 that the RMF theory with force parameters NL-SH and TM2 provides a good description on the properties of He isotopes. As a conclusion, the RMF theory also predicts a one-neutron halo in $^5$He.

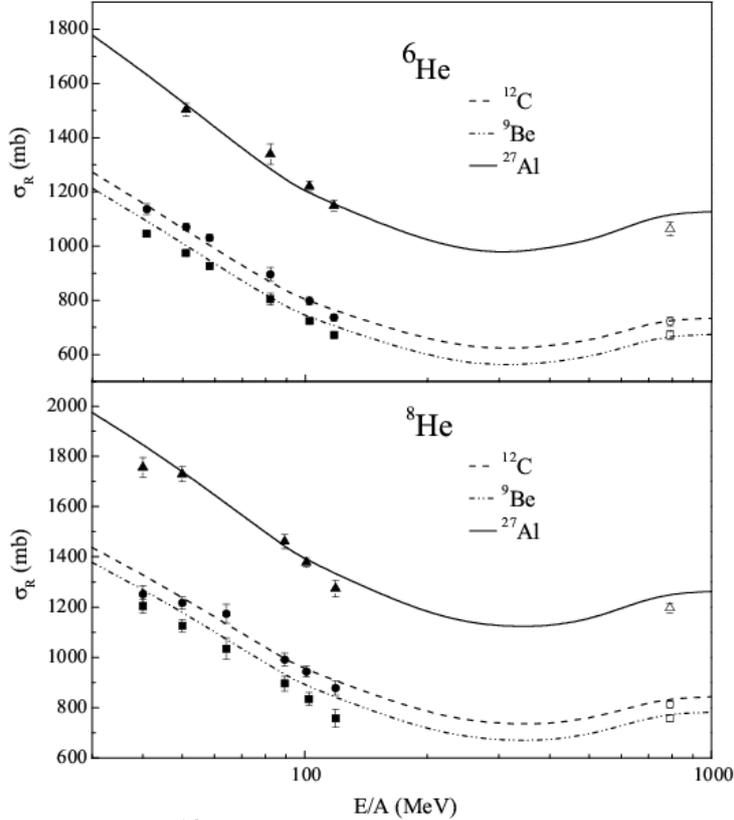

Fig. 2. The reaction cross sections of $^{6,8}$He as a function of beam energy. The open symbols denote data from Refs. [28]. The closed symbols denote data from Refs. [26]. The curves are calculated by use of the Glauber model in conjunction with the RMF theory with force parameter of NL-SH.

In conclusion, we have calculated the ground state properties of He isotopes using the RMF theory with NL-SH and TM2 force parameters. It is shown that neutrons in $1p_{3/2}$ state in He isotopes are weakly bound. The $^5$He exhibits a halo structure, and the size of neutron halos in it is larger than that in $^6$He. The introduction of the Glauber model confirms the reliabilities of the investigations of the RMF theory. At last, we believe that the many-body space information of $^{5,6,7}$He, are reliable.

**Acknowledgements**

We would like to acknowledge the financial support provided by Anhui University of Science and Technology (11130).